\documentclass[%
 reprint,
 amsmath,amssymb,
 aps,
 pra
]{revtex4-2}

\usepackage{graphicx}
\usepackage{dcolumn}
\usepackage{bm}
\usepackage{mathtools}
\usepackage{ulem}
\mathtoolsset{showonlyrefs,showmanualtags}
\usepackage{color}
\usepackage{comment}
\usepackage{xspace}

\newcommand{\dir}{X}
\newcommand{\obs}{Y}
\newcommand{\Post}{Z}
\newcommand{\post}{z}
\newcommand{\W}{W}

\newcommand{\mi}{p}
\newcommand{\pred}{\mathrm{pred}}
\newcommand{\sen}{\mathrm{obs}}

\newcommand{\td}{\theta}

\newcommand{\rr}{\epsilon}

\newcommand{\ttheta}{\Theta}

\newcommand{\uni}{\bm{\mathrm{e}}}

\newcommand{\dc}{\nabla c}
\newcommand{\tilt}{\tilde{t}}

\newcommand{\argmin}{\mathop{\rm arg~min}\limits}
\newcommand{\dif}{\mathrm{d}}
\newcommand{\dd}{\mathcal{D}}

\newcommand{\Prob}{\mathbb{P}}

\newcommand{\Expect}{\mathbb{E}}

\newcommand{\tp}{\mathbb{T}}

\newcommand{\rbra}[1]{\left(#1\right)}
\newcommand{\cbra}[1]{\left\{#1\right\}}
\newcommand{\sbra}[1]{\left[#1\right]}
\newcommand{\abra}[1]{\left\langle#1\right\rangle}

\newcommand{\e}{\mathrm{e}}
\newcommand{\pf}[2]{\frac{\partial #1}{\partial #2}}
\newcommand{\ff}[2]{\frac{\delta #1}{\delta #2}}

\newcommand{\fss}[3]{\frac{\delta^{2} #1}{\delta #2 \delta #3}}

\newcommand{\ps}[2]{\frac{\partial^2 #1}{\partial #2^2}}

\newcommand{\dpro}[2]{\abra{#1,#2}}

\newcommand{\ii}{\mathrm{i}}

\newcommand{\reserve}[1]{}

\newcommand{\eqnref}[1]{Eq. \eqref{#1}}

\begin{document}

\preprint{APS/123-QED}

\title{Gradient sensing limit of a cell when controlling the elongating direction}

\author{Kento Nakamura}
\affiliation{%
RIKEN Center for Brain Science, 2-1 Hirosawa, Wako, Saitama 351-0198, Japan
}%
\author{Tetsuya J. Kobayashi}%
\email{tetsuya@mail.crmind.net}
\affiliation{Institute of Industrial Science, The University of Tokyo, 4-6-1, Komaba, Meguro-ku, Tokyo 153-8505 Japan}
\altaffiliation[Also at ]{Universal Biology Institute, The University of Tokyo, 7-3-1, Hongo, Bunkyo-ku, Tokyo 113-8654, Japan.}
\altaffiliation[Also at ]{%
$^1$Department of Mathematical Informatics, the Graduate School of Information Science and Technology, the University of Tokyo, 7-3-1, Hongo, Bunkyo-ku, Tokyo, 113-8654, Japan
}%




\date{\today}

\begin{abstract}
Eukaryotic cells perform chemotaxis by determining the direction of chemical gradients based on stochastic sensing of concentrations at the cell surface.
To examine the efficiency of this process, previous studies have investigated the limit of estimation accuracy for gradients.
However, these studies assume that the cell shape and gradient are constant, and do not consider how adaptive regulation of cell shape affects the estimation limit.
Dynamics of cell shape during gradient sensing is biologically ubiquitous and can influence the estimation by altering the way the concentration is measured, and cells may strategically regulate their shape to improve estimation accuracy.
To address this gap, we investigate the estimation limits in dynamic situations where cells change shape adaptively depending on the sensed signal.
We approach this problem by analyzing the stationary solution of the Bayesian nonlinear filtering equation.
By applying diffusion approximation to the ligand-receptor binding process and the Laplace method for the posterior expectation under a high signal-to-noise ratio regime, we obtain an analytical expression for the estimation limit.
This expression indicates that estimation accuracy can be improved by elongating perpendicular to the estimated direction, which is also confirmed by numerical simulations.
Our analysis provides a basis for clarifying the interplay between estimation and control in gradient sensing and sheds light on how cells optimize their shape to enhance chemotactic efficiency.
\end{abstract}

\maketitle

\section{Introduction}
\label{Introduction}
Cells have evolved sophisticated mechanisms to sense and adapt to their fluctuating environment. An example of such adaptation is eukaryotic chemotaxis, manifested by the directed movement of cells along chemical gradients  \cite{parent1999cell}. In eukaryotic cells, chemotaxis relies on the spatial sensing of chemical gradients using an array of receptors distributed across the cell surface. The reliable detection of chemical gradients is challenging due to the inherent stochasticity of receptor-ligand binding \cite{ueda2001single, ueda2007stochastic}. This stochasticity introduces measurement noise in the local ligand concentration and leads to errors in estimating spatial difference of the concentrations. Despite this uncertainty, eukaryotic cells exhibit a remarkable ability to detect and navigate shallow gradients \cite{song2006dictyostelium, van2007biased, fuller2010external,amselem2012control}, raising the question of how sensory systems are optimized to process noisy chemical information. To address this question, theoretical studies have sought to elucidate the limits on the accuracy of chemical gradient sensing in eukaryotic cells set by stochastic uncertainty \cite{ueda2007stochastic, endres2008accuracy, rappel2008receptor, hu2010physical, hu2011quantifying,aquino2014memory,novak2021bayesian}. These studies have employed tools from estimation theory, information theory, and statistical physics to derive bounds on the precision of gradient sensing.

One notable factor that can influence the accuracy of gradient sensing is the shape of the cell. During chemotaxis, eukaryotic cells undergo significant shape changes, such as elongation, pseudopod extension, and polarization \cite{tweedy2013distinct,van2017coupled, imoto2021comparative}. These shape changes modulate the spatial distribution of receptors on the cell surface, altering the way in which the cell samples the chemical gradients. For instance, an elongated cell may be more sensitive to concentration differences along its long axis but less sensitive to differences along its short axis. By dynamically regulating its shape, a eukaryotic cell may enhance its gradient sensing performance by sampling gradient information more effectively.

However, the theoretical understanding of how cell shape affects gradient sensing remains limited. Previous studies have investigated the impact of cell elongation on the static gradient sensing based on the optimal estimation theory \cite{hu2011geometry, baba2012directional}. These studies have revealed that the estimation accuracy of gradient depends on the spatial configuration between the cell and the true gradient, such as the angle between the cell’s elongation axis and the gradient direction. However, these studies assume that the elongation direction is temporally fixed and externally determined, not actively adjusted by the cell in response to its environment. This assumption limits the applicability of their findings to actual scenarios where cells continuously adjust their elongation direction based on sensory inputs. Another line of studies has addressed this limitation by developing a phenomenological model that couples cell shape dynamics with an internal polarity mechanism that represents the cell's estimate of the gradient direction \cite{hiraiwa2013theoretical,tweedy2013distinct}. While these models provide valuable insights into the interplay between shape and sensing during chemotaxis, it does not provide a framework for deriving estimation limit attained by dynamic cell shape control.

In this work, we bridge this gap by developing optimality theory to derive the gradient sensing limit under the feedback control of cell shape. We frame the gradient sensing problem as a Bayesian nonlinear filtering problem \cite{bain2009fundamentals,kutschireiter2020hitchhiker}, where the cell estimates the gradient direction from noisy receptor activity measurements while controlling its shape. The Bayesian nonlinear filtering theory has been applied to derive the optimal cellular sensing strategy under fluctuating environments \cite{kobayashi2010implementation, zechner2016molecular, mora2019physical,nakamura2021connection}. The theory has recently been applied to the gradient sensing by a eukaryotic cell with a circular shape \cite{novak2021bayesian,auconi2022gradient}, but has not been used under feedback control of sensory apparatus. We show that the Bayesian nonlinear filtering theory can be combined with feedback control strategy of sensory processes.

Our paper is organized as follows. In Section II, we introduce a mathematical model of gradient sensing by a cell with a dynamically controlled shape. This model incorporates the stochastic dynamics of receptor-ligand binding and the feedback control of cell shape based on the sensory information. In Section III, we formulate the gradient sensing problem as a Bayesian nonlinear filtering problem and derive the optimal filtering equation for the posterior distribution of the gradient direction, which sets a limit on the accuracy of gradient sensing.  In Section IV, we specialize our analysis to the case where the cell can control its elongation direction to be either parallel or perpendicular to the estimated gradient direction. In this case, we derive an analytical expression for the asymptotic estimation error in the limit of high signal-to-noise ratio, and we validate our theory using numerical simulations. Finally, in section V, we conclude this work by discussing the possible directions for future research.

\section{Model}
\label{Model}
\subsection{Setting}
\begin{figure*}
    \centering
    \includegraphics[width=\linewidth]{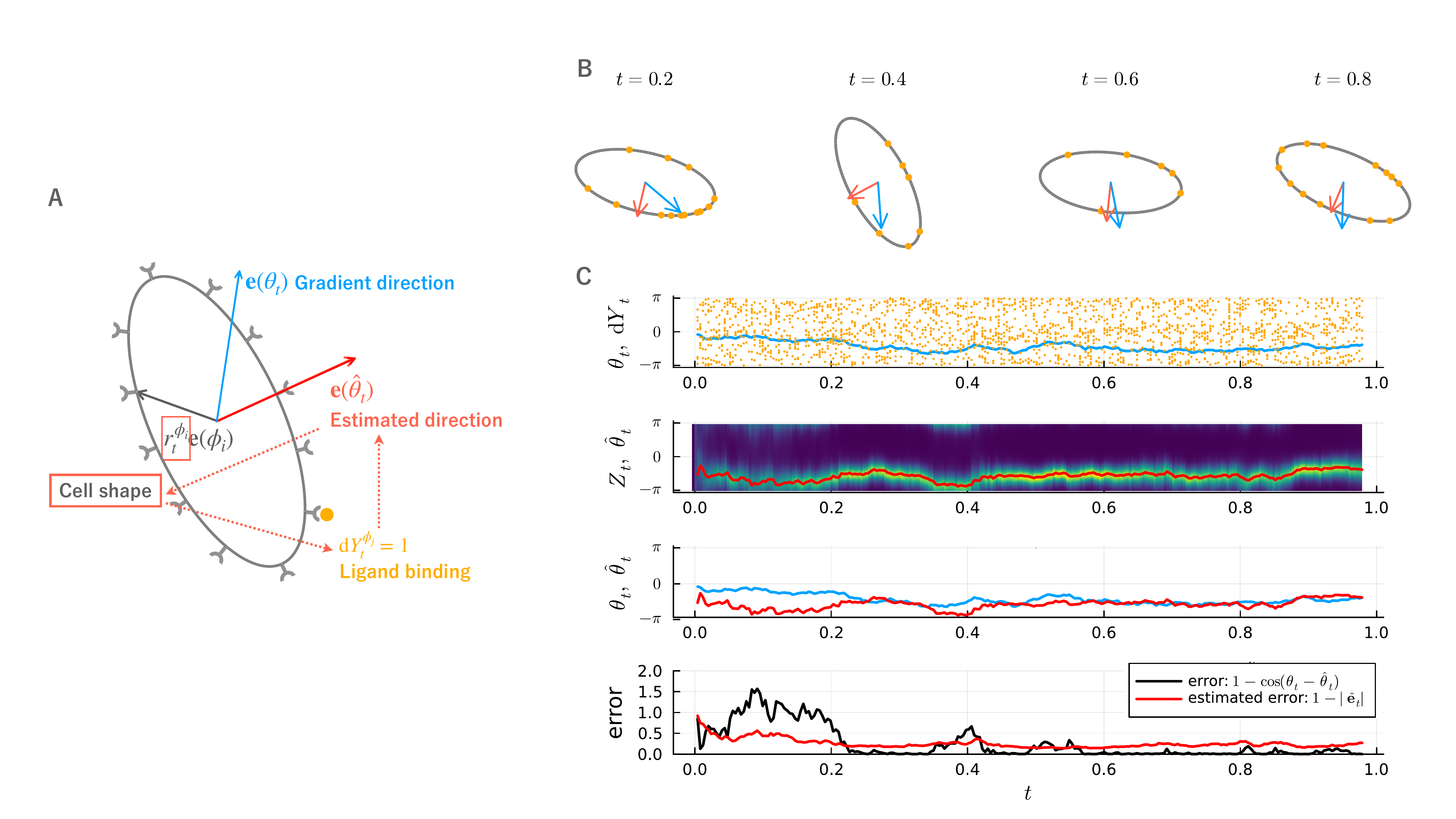}
    \caption{
    (A)Schematic diagram of gradient sensing under feedback control of cell shape.
    The cell shape is controlled based on the estimate of the gradient direction and, in turn, affects the gradient estimation by biasing the ligand binding process.
    (B)Time lapse of a sample process of the gradient sensing.
    Blue and red arrows indicate the true and estimated gradient direction, respectively.
    The length of the red arrows indicate the confidence of estimate $|\hat\uni_{t}|$.
    Orange dots show the locations of receptors at which ligand bound during time $[t,t+\Delta t)$ with $\Delta t$ being the discretization width.
    (C)Time evolution of variables in the sample process shown in (B).
    The first row shows the true gradient direction $\theta_{t}$ (blue curve) and the angular locations of receptors at which ligand binds $\dif\obs_{t}$ (orange dots).
    The second row shows the posterior density of the gradient direction $\Post_{t}$ and the estimated gradient direction $\hat\theta_{t}$ (red line).
    The third row shows the true (blue curve) and estimated (red curve) gradient direction.
    The final row shows the error between the true and estimated direction measured by $(\theta_{t}-\hat\theta_{t})^{2}/2$ (black curve) and the uncertainty of estimate represented by $1-|\hat\uni_{t}|$ (red curve).
    The trajectories were simulated under the signal-to-noise ratio $\Lambda = 10^1$ with the following dimensionless parameter values: $\lambda = 2\cdot 10^3, \alpha = 10^{-1}$, and $\sigma\epsilon = -0.4$.
    }
    \label{fig:schematic}
\end{figure*}

We formulate a spatial gradient sensing model that incorporates cell-shape modulation (Figure \ref{fig:schematic}).

We consider a cell placed at the origin of a 2-dimensional space with a linear concentration gradient.
The gradient direction is denoted by $\td_{t}\in[0,2\pi)$ and is assumed to fluctuate over time following a circular diffusion process
\begin{align}
    \dif\td_{t} = \sqrt{2D}\dif\W_{t}, \label{eq:Direction/motion}
\end{align}
where $\W_{t}$ is a standard Wiener process and $D$ is the rotational diffusion coefficient.
The ligand concentration $c(\bm{x},\td)$ at a given position $\bm{x}\in\mathbb{R}^{2}$ when the gradient direction is $\td_{t} = \td$ is given by
\begin{align}
    c(\bm{x},\td) = c_{0}+\dc\abra{\bm{x}, \uni(\td)}.
\end{align}
Here, $c_{0}$ and $\dc$ are constants representing the concentration at the origin and the gradient steepness, respectively, and $\uni(\theta) = [\cos(\theta),\sin(\theta)]^{\tp}$ is a unit vector pointing in the direction $\theta$.

The cell senses the gradient using receptors on its surface, represented as a closed curve $r_{t} = \{r_{t}^{\phi}\mid \phi\in[0,2\pi)\}$ based on the Fourier series:
\begin{align}
    r_{t}^{\phi} = r_{0}\rbra{1 + \sum_{k=2}^{\infty}\rr_{t}^{k}\cos(k(\phi-\phi_{t}^{k}))}
\end{align}
where $\rr_{t}^{k}\in\mathbb{R}$ and $\phi_{t}^{k}\in[0,2\pi)$ are the amplitude and phase of the $k$-th Fourier component, respectively.
The first Fourier component is removed to ensure that the cell center $\oint\dd\phi r_{t}^{\phi}$ coincides with the coordinate origin.
Given the cell radius $r_{t}^{\phi}$, we assume that $N\in\mathbb{N}$ receptors are located on the cell surface at equal arc length intervals.
The direction of the $i$-th receptor is denoted by $\phi_{i}$, and the ligand binds to the receptor with a rate proportional to the local ligand concentration $c(r_{t}^{\phi_{i}}\uni(\phi_{i}) ,\td)$ at the receptor's position $r_{t}^{\phi_{i}}\uni(\phi_{i})$.
We assume that the unbinding time is negligible compared to the time binding time.
The number of ligand binding events at the $i$-th receptor up to time $t$ is described by the Poisson process $Y_{t}^{\phi_{i}}$ with rate process
\begin{align}
    \lambda_{t}^{\phi_{i}} = \lambda^{\phi_{i}}(r_{t},\theta_{t}),\ \lambda^{\phi_{i}}(r,\theta) &:= \frac{\lambda_{0}}{N}c(r^{\phi_{i}}\uni(\phi_{i}) ,\theta).
\end{align}
where $\lambda_{0}/N$ is the binding rate of a single receptor under unit concentration.

Finally, we assume that the cell controls its shape $r_{t}$ based on the ligand binding history measured by all receptors.
By introducing the combined notation for the ligand binding process over receptors, $Y_{t} := \{Y_{t}^{\phi_{i}}\mid i = 1,\ldots,N\}$, the feedback control assumption is expressed by setting $\epsilon_{t}^{k}$ and $\phi_{t}^{k}$ as functionals of the history of $Y_{t}$:
\begin{align}
    \epsilon_{t}^{k} = \epsilon_{t}^{k}[Y_{0:t^{-}}],\ \phi_{t}^{k} = \phi_{t}^{k}[Y_{0:t^{-}}].
\end{align}
Here, $Y_{0:t^{-}} = \{Y_{s}\mid s\in[0,t)\}$ represents the history of observation process up to, but not including, time $t$.
The exclusion of $Y_{t}$ ensures causality between sensing and control; otherwise, the value of $Y$ after a jump at time $t$ would affect whether the jump occurs at that time.
In Section \ref{Derivation}, we will consider a more concrete setting for feedback control of the cell shape to derive the analytical result.

\subsection{Dimensionless system}
To narrow down the parameter space without losing generality, we consider a dimensionless system on the above setting.
By rescaling time as $\tilt:= t/\tau_{D}$ with $\tau_{D}:=1/D$, we obtain the dimensionless dynamics of the gradient direction:
\begin{align}
    \dif\theta_{\tilt} &= \sqrt{2}\dif W_{\tilt}, \label{eq:model_dimless_theta}
\end{align}
By introducing several dimensionless parameters, the ligand-receptor binding process is obtained as the Poisson process $Y_{\tilde{t}}^{\phi_{i}}$ with the following dimensionless rate process:
\begin{align}
    \tilde\lambda_{\tilt}^{\phi_{i}}(\theta)&:=\lambda_{\tilt\tau_{D}}^{\phi_{i}}(\theta)\tau_{D} = \frac{\bar\lambda}{N}\rbra{1 + \alpha \tilde{r}_{\tilt}^{\phi_{i}}\abra{\uni(\phi_{i}),\uni(\theta)}}. \label{eq:model_dimless_rate}
\end{align}
Here, $\tilde{r}_{\tilt}^{\phi_{i}}:=r_{\tilt\tau_{D}}^{\phi_{i}}/r_{0}$ is the normalized cell radius, and $\bar\lambda:=\lambda_{0}c_{0}\tau_{D}$ and $\alpha:=r_{0}\dc/c_{0}$ are dimensionless parameters. $\bar\lambda$ represents the binding number during the diffusion time $\tau_{D}$ in the absence of a gradient, and $\alpha$ represents the relative difference in ligand concentration across the cell radius.

In the following sections, we consider the dimensionless system and omit the tildes from $\tilde{t},\tilde{\lambda},$ and $\tilde{r}$ for visual clarity.

\section{Estimation and control by Bayesian nonlinear filtering}
\label{Filtering}
We formulate a sensing problem for the gradient direction $\theta_{t}$ under the feedback control of the cell shape $r_{t}$, and describe the optimal estimator derived based on the Bayesian nonlinear filtering theory.
We assume that the cell estimates the gradient direction $\theta_{t}$ based on the sensing history $Y_{0:t}:=\{Y_{s}\mid s\in[0,t]\}$, i.e., the estimated direction is expressed as $\vartheta_{t}[\obs_{0:t}]$ with a functional $\vartheta_{t}$.

The goal is to find the functional $\vartheta_{t}$ that optimizes an estimation performance.
As a measure of estimation performance for directional statistics, we consider the circular variance of the difference between the true and estimated gradient direction, as in the previous study \cite{novak2021bayesian}:
\begin{align}
    \mathrm{CV}(\theta_{t}-\vartheta_{t}[\obs_{0:t}]) := 1-\left|\Expect^{\theta_{t},\obs_{0:t}}\sbra{\exp\rbra{\mathrm{i}(\theta_{t}-\vartheta_{t}[\obs_{0:t}])}}\right|,\label{eq:circular variance}
\end{align}
where $\mathrm{i}$ is an imaginary unit and $\Expect^{\theta_{t},\obs_{0:t}}$ represents the expectation with respect to the joint distribution of $\theta_{t}$ and $\obs_{0:t}$.
As described in Ref. \cite{novak2021bayesian}, the circular variance relates to the square error as $\mathrm{CV}(\theta_{t}-\vartheta_{t}[\obs_{0:t}])\approx \Expect[(\theta_{t}-\vartheta_{t}[\obs_{0:t}])^{2}]/2$ when the estimated direction concentrates around the true gradient direction.
The optimal estimator $\hat\theta_{t}$ that minimizes the estimation error is represented based on the posterior density function $\Post_{t}$ of the gradient direction $\theta_{t}$ conditioned on the sensing history $Y_{0:t}$ (see Appendix \ref{Appendix/Estimation_limit} for derivation):
\begin{align}
    \hat\theta_{t} &= \vartheta_{t}^{\ast}[\obs_{0:t}]:=\argmin_{\vartheta_{t}[\obs_{0:t}]}\mathrm{CV}(\theta_{t}-\vartheta_{t}[\obs_{0:t}]) = \arg\hat\uni_{t}, \label{eq:estimation_optimal_estimator}\\
    \hat\uni_{t}&:=\Expect^{\theta_{t}}[\uni(\theta_{t})\mid \obs_{0:t}] = \oint\dd\theta\uni(\theta)\Post_{t}(\theta), \label{eq:estimation_vector}
\end{align}
where $\Post_{t}(\theta) := \Prob\rbra{\theta_{t}=\theta\mid \obs_{0:t}}$ is the posterior density function of $\theta_{t}$ given $\obs_{0:t}$, and $\Expect^{\theta_{t}}[\cdot\mid\obs_{0:t}]$ represents the expectation with respect to $\Post_{t}(\theta)$.
$\hat\uni_{t}$ is a vector whose argument represents the estimated direction $\hat\theta_{t}$, and its magnitude $\|\hat\uni_{t}\|\leq 1$, where $\|\cdot\|$ denotes a Euclidean norm, represents the certainty of the estimate.
The magnitude takes the minimum $\|\hat\uni_{t}\|= 0$ when the posterior density function is uniform, $\Post_{t}(\theta) = (2\pi)^{-1}$, and the maximum $\|\hat\uni_{t}\|=1$ when the posterior density concentrates completely on the estimated direction, $\Post_{t}(\theta) = \delta(\theta-\hat\theta_{t})$.

Using the vector $\hat\uni_{t}$, we can express the limit of estimation error attained by the optimal estimator:
\begin{align}
    \min_{\vartheta_{t}}\mathrm{CV}(\theta_{t}-\vartheta_{t}[\obs_{0:t}]) = 1-\Expect^{Y_{0:t}}[|\hat\uni_{t}|],\label{eq:limit_min}
\end{align}
where $\Expect^{Y_{0:t}}$ represents the expectation with respect to the marginal distribution of $Y_{0:t}$ with $\theta_{t}$ being averaged out.

To calculate the optimal estimator based on the posterior density function $\Post_{t}(\theta)$, we can use the following stochastic partial differential equation known as the Kushner equation in the nonlinear filtering theory (for a pedagogical introduction, see Ref. \cite{kutschireiter2020hitchhiker}; for detailed derivations, see the references therein and Ref. \cite{venugopal2015ensemble}):
\begin{align}
    \dif\Post_{t}(\theta)
    &= \dif^{\pred}\Post_{t}(\theta) + \dif^{\sen}\Post_{t}(\theta),\label{eq:filter}\\
    \dif^{\pred}\Post_{t}(\theta)&=\ps{Z_{t}}{\theta}(\theta)\dif t,\label{eq:filter_pred}
\end{align}
\begin{align}
    \dif^{\sen}\Post_{t}(\theta)=\Post_{t^{-}}(\theta)\sum_{i=1}^{N}\frac{\lambda_{t^{-}}^{\phi_{i}}(\theta)-\hat\lambda_{t^{-}}^{\phi_{i}}}{\hat\lambda_{t^{-}}^{\phi_{i}}}\rbra{\dif\obs_{t}^{\phi_{i}}-\hat\lambda_{t^{-}}^{\phi_{i}}\dif t}.\label{eq:filter_obs}
\end{align}
Here, $\Post_{t^{-}}$ and $\hat\lambda_{t^{-}}^{\phi_{i}}:=\Expect^{\theta_{t}}[\lambda_{t^{-}}^{\phi_{i}}(\theta_{t^{-}})\mid\obs_{0:t^{-}}]$ with $t^{-}$ representing the time infinitesimally before $t$ are statistics conditioned on $\obs_{0:t^{-}}$. This conditioning ensures that the coefficient of the jump term $\dif\obs_t^{\phi_i}$ in the equation is independent of the value of $\obs$ at time $t$.
The first term, $\dif^{\pred}\Post_{t}(\theta)$, represents the prediction based on the prior knowledge about the dynamics of $\theta_{t}$ and the second term, $\dif^{\sen}\Post_{t}(\theta)$, represents the correction based on the sensed signal $\dif\obs_{t}$.

\section{Analytical derivation of estimation limit}
\label{Derivation}
In this section, we derive an analytical expression for the estimation limit by approximating the stationary solution of the filtering equation.
We introduce additional assumptions to make the problem analytically tractable.
\subsection{Additional assumptions for an analytical treatment}
\begin{figure*}
    \centering
    \includegraphics[width=\linewidth]{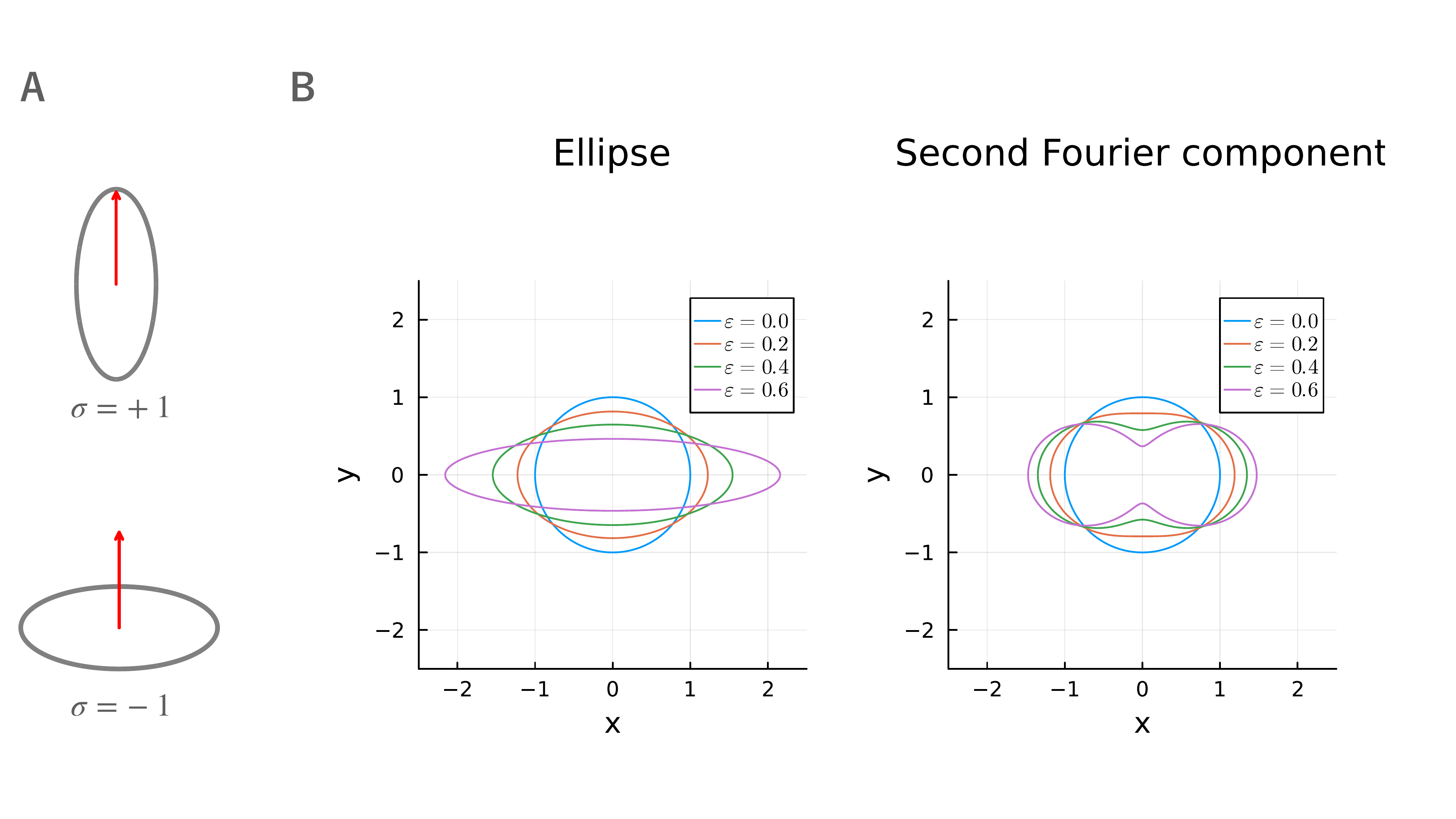}
    \caption{Parameters of controlled cell elongation. (A) $\sigma$ dependency of the relative angle between elongation direction (gray curve indicates an elongating cell) and the estimated direction (red arrow). A cell elongates along the estimated direction if $\sigma=+1$ (upper panel) and elongates perpendicularly to the estimated direction if $\sigma=-1$ (lower panel). (B) $\epsilon$ dependency of the extent of elongation for an elliptic shape (left) and a shape described by the second Fourier component (right).}
    \label{fig:elongation_pattern}
\end{figure*}
To derive the estimation limit analytically, we focus on the control pattern of cell shape such that the cell elongates either parallel or perpendicular to the estimated direction $\hat\theta_{t}$ (Figure \ref{fig:elongation_pattern}A).
We model this control pattern by imposing a condition on the second mode of the cell shape:
\begin{align}
    \epsilon_{t,2}[\obs_{0:t^{-}}] = \sigma\epsilon,\ \phi_{t,2}[\obs_{0:t^{-}}] = \hat\theta_{t^{-}}=\vartheta_{t^{-}}^{\ast}[\obs_{0:t^{-}}],
\end{align}
where $\epsilon \geq 0$ and $\sigma\in\{\pm 1\}$ are constants.
When $\sigma=+1$ and $\sigma=-1$, the cell elongates parallel and perpendicular to the estimated direction, respectively (Figure \ref{fig:elongation_pattern}A).
In addition, we consider the first order perturbation of the cell shape around a circular one by assuming $\epsilon_{t,k}=O(\epsilon)$ for every $k$ and ignoring the second and higher order terms, $O(\epsilon^{2})$.

We further restrict the parameter region of interest for analytical tractability.
We assume that the ligand gradient is shallow so that the difference in ligand concentration across the spatial scale of the cell $\alpha r_{t}^{\phi_{i}}$ is sufficiently small.
Second, we assume that there are a sufficiently large number of receptors, $N\gg 1$.
Finally, we assume that $\Lambda := \bar{\lambda}\alpha^{2}/2$ is sufficiently large.
This assumption indicates that the cell is provided with sufficient amount of gradient information because the normalized ligand binding rate $\bar{\lambda}$ and gradient steepness $\alpha$ are high.
We note that $\Lambda$ is a dimensionless version of the ``information rate" defined in the previous study \cite{novak2021bayesian}.

\subsection{Derivation of estimation limit}
The filtering equation is generally difficult to analyze because the posterior density function $\Post_{t}(\theta)$ is not closed as a finite-dimensional model except in some special cases.
To simplify the problem, we first discuss the behavior of the observation term and then combine the effect of the prediction term.
\subsubsection{Coordinate transform of the observation term}
We can see from \eqnref{eq:filter_obs} that the observation term is proportional to $\Post_{t}(\theta)$ except for $\hat{\lambda}_{t}^{\phi_{i}}$, suggesting that $\Post_{t}(\theta)$ may grow exponentially.
For analyzing an exponentially growing variable, we employ the coordinate transformation to the logarithm of the original variable.

To easily transform the dynamics of $\Post_{t}(\theta)$ to that of the log-posterior density $\log\Post_{t}(\theta)$, we apply a diffusion approximation of $\dif\obs_{t}^{\phi_{i}}$ in advance based on the weak gradient assumption.
By ignoring $o((\alpha r_{t}^{\phi_{i}})^{2})$ terms in the corresponding Fokker-Planck equation, we obtain:
\begin{align}
    \dif^{\sen}\Post_{t}(\theta)
    &\approx \Post_{t}(\theta)
    \sum_{i=1}^{N}\sqrt{\frac{2\Lambda}{N}} r_{t}^{\phi_{i}}\abra{\uni(\phi_
{i}),\uni(\theta)-\hat{\uni}_{t}}\dif\W_{t}^{i},\quad\label{eq:dz_diffusion}
\end{align}
where $\dif\W_{t}^{i}$ with $i=1,\ldots,N$ are the mutually independent standard Wiener processes (see Appendix \ref{Appendix/Filtering_general} for derivation).
Here, we have also taken the expectation over the true gradient direction $\theta_{t}$ because we only need the statistics of $\hat\uni_{t}$ marginalized over $\theta_{t}$ to obtain the estimation limit (\eqnref{eq:limit_min}).

By applying the Ito's formula to Eq. \eqref{eq:dz_diffusion}, we get the dynamics of the log-posterior density:
\begin{widetext}
\begin{align}
    \dif^{\sen}\log\Post_{t}(\theta) 
    &\approx \Post_{t}(\theta)^{-1}\dif^{\sen}\Post_{t}(\theta)-\frac{1}{2}\Post_{t}(\theta)^{-2}\dif^{\sen}[\Post(\theta)]_{t}\\
    &\approx  \sum_{i=1}^{N}\sqrt{\frac{2\Lambda}{N}} r_{t}^{\phi_{i}}\abra{\uni(\phi_{i}),\uni(\theta)-\hat{\uni}_{t}}\dif\W_{t}^{i}
    - \sum_{i=1}^{N}\frac{\Lambda}{N}(r_{t}^{\phi_{i}})^{2}\abra{\uni(\phi_{i}),\uni(\theta)-\hat{\uni}_{t}}^{2}\dif t\\
    &= - \sum_{i=1}^{N}\frac{\Lambda}{N}(r_{t}^{\phi_{i}})^{2}\abra{\uni(\phi_{i}),\uni(\theta)-\hat{\uni}_{t}}^{2}\dif t + O(\Lambda^{1/2}),
    \label{eq:obs_dlogZ}
\end{align}
where $\dif^{\sen}[\Post(\theta)]_{t}$ is the quadratic variation process.
We neglected the fluctuating part of the observation term (the first term on the second line of Eq. \eqref{eq:obs_dlogZ}) based on the large $\Lambda$ assumption.
We note that the variance of the fluctuating part does not diverge in the limit of $N\rightarrow \infty$ because $\dif\W_{t}^{i}$ is independent for each $i$.

When viewed in the logarithmic coordinate (Eq. \eqref{eq:obs_dlogZ}), $\Post_{t}(\theta)$ dependent factors disappear from the observation term.
The dependence on $\theta$ is now simply represented by $\abra{\uni(\phi_{i}),\uni(\theta)-\hat{\uni}_{t}}^{2}$, which is further expanded up to a $\theta$-independent constant:
\begin{align}
    \abra{\uni(\phi_{i}),\uni(\theta)-\hat{\uni}_{t}}^{2}
    &=\rbra{\abra{\uni(\phi_{i}),\uni(\theta)}-|\hat\uni_{t}|\abra{\uni(\phi_{i}),\uni(\hat\theta_{t})}}^2\\
    &= \frac{1}{2}\abra{\uni(2\phi_{i}),\uni(2\theta)}+|\hat\uni_{t}|\rbra{\abra{\uni(\hat\theta_{t}),\uni(\theta)} - \abra{\uni(2\phi_{i}),\uni(\theta+\hat\theta_{t})}} + const. \\
    &= \frac{1}{2}\abra{\uni(2(\phi_{i}-\hat\theta_{t})),\uni(2(\theta-\hat\theta_{t}))}-|\hat\uni_{t}|\rbra{\abra{\uni(0)+\uni(2(\phi_{i}-\hat\theta_{t})),\uni(\theta-\hat\theta_{t})}} + const. \label{eq:inner_product_likelihood}
\end{align}

\subsubsection{Cell shape dependence of the observation term}
We can summarize the dependence of the observation term on cell shape by substituting Eq. \eqref{eq:inner_product_likelihood} into Eq. \eqref{eq:obs_dlogZ} and taking the sum over $i$:
\begin{align}
    \dif^{\sen}\log\Post_{t}(\theta) &\approx -\Lambda\rbra{ \frac{1}{2}\abra{\bm{R}_{2},\uni(2(\theta-\hat\theta_{t}))}-|\hat\uni_{t}|\abra{\bm{R}_{0}+\bm{R}_{2},\uni(\theta-\hat\theta_{t})} }\dif t+O(\Lambda^{1/2}).\label{eq:obs_dlogZ_2}
\end{align}
Here, $\bm{R}_{k} := N^{-1}\sum_{i=1}^{N}(r_{t}^{\phi_{i}})^{2}\uni(k(\phi_{i}-\hat\theta_{t}))$ is a quantity that summarizes the cell shape and its configuration with respect to the estimated direction  $\hat\theta_{t}$.
Specifically, $\bm{R}_{0}$ and $\bm{R}_{2}$ represent the overall radius and elongation of the cell, respectively.
In the limit of large $N$ and small $\epsilon$,  $\bm{R}_{0}$ and $\bm{R}_{2}$ are approximated as follows:
\begin{align}
    \bm{R}_{0}&\approx \rbra{\oint\dd\phi\rho_{t}^{\phi}}^{-1}\oint \dd\phi\rho_{t}^{\phi}(r_{t}^{\phi})^{2}\uni(0) \approx \uni(0) + o(\epsilon),\label{eq:R0}\\
    \bm{R}_{2}&\approx \rbra{\oint\dd\phi\rho_{t}^{\phi}}^{-1}\oint \dd\phi\rho_{t}^{\phi}(r_{t}^{\phi})^{2}\uni(2(\phi-\hat\theta_{t})) \approx \frac{3}{2}\sigma\epsilon\uni(0) + o(\epsilon).\label{eq:R2}
\end{align}
Here, we defined $\rho_{t}^{\phi}$ such that $\rho_{t}^{\phi}\dd\phi$ indicates the arc length of the cell circumference spanned by the angle interval $[\phi,\phi+\dd\phi)$ and used the approximation $\rho_{t}^{\phi}(r_{t}^{\phi})^{2}\approx 1 + 3(r_{t}^{\phi}-1)$, which holds for sufficiently small $\epsilon$.
The term $\rho_{t}^{\phi}$ is introduced due to the assumption that receptors are distributed equally with respect to arc length intervals, rather than angular intervals.
By substituting Eqs. \eqref{eq:R0} and \eqref{eq:R2} into Eq. \eqref{eq:obs_dlogZ_2}, we get the following approximation of the observation term:
\begin{align}
    \dif^{\sen}\log\Post_{t}(\theta) &\approx
     -\Lambda\rbra{\frac{3}{4}\sigma\epsilon\cos(2(\theta-\hat\theta_{t})) - |\hat\uni_{t}|\rbra{1 + \frac{3}{2}\sigma\epsilon}\cos(\theta-\hat\theta_{t})}\dif t + O(\Lambda^{1/2})\\
     &= \Lambda\rbra{|\hat\uni_{t}|\cos(\theta-\hat\theta_{t}) + \frac{3}{2}\sigma\epsilon \rbra{|\hat\uni_{t}|\cos(\theta-\hat\theta_{t})-\frac{1}{2}\cos(2(\theta-\hat\theta_{t}))}}\dif t + O(\Lambda^{1/2}).
    \label{eq:obs_fourier_final}
\end{align}
Eq. \eqref{eq:obs_fourier_final} indicates that, when $\epsilon$ is sufficiently small, the cell shape affects the estimation through its second Fourier mode $\sigma\epsilon$ and that $\Post_{t}$ would concentrate around the estimated direction $\hat\theta_{t}$ under large $\Lambda$.
\end{widetext}

.

\subsubsection{Integration of
the effect of the prediction term}
We incorporate the effect of the prediction term by using the concentration property of $\Post_{t}(\theta)$ around $\hat\theta_{t}$ suggested by Eq. \eqref{eq:obs_fourier_final}.
By assuming that the posterior density is unimodal with a maximum point $\theta_{t}^{\max}$, which eventually coincides with $\hat\theta_{t}$, we expand $\log\Post_{t}$ into a quadratic function:
\begin{align}
    \log\Post_{t}(\theta) \approx -\frac{1}{2}\kappa_{t}(\theta-\theta_{t}^{\max})^{2} + o((\theta-\theta_{t}^{\max})^{2}) + \mathrm{const.}
\end{align}
where $\kappa_{t} := -\ps{}{\theta}\log\Post_{t}(\theta_{t}^{\max})>0$ indicates the concentration of the posterior density.
When $\Lambda$ is high, the posterior density would concentrate on a maximal point and $\kappa_{t}$ would be large.
For a large value of $\kappa_{t}$, we can use Laplace's method to approximate the posterior expectation:
\begin{align}
    \Expect\sbra{F(\theta_{t})\mid \obs_{0:t}} 
    &= \oint\dd\theta F(\theta)\exp(\log\Post_{t}(\theta))\\
    &\approx\oint \dd\theta F(\theta)\exp\rbra{-\frac{1}{2}\kappa_{t}(\theta-\theta_{t}^{\max})^{2}}\\
    &\approx F(\theta_{t}^{\max}) + \frac{1}{2}F''(\theta_{t}^{\max})\kappa_{t}^{-1} + o(\kappa_{t}^{-1}).
\end{align}
By substituting $F(\theta) = \uni(\theta)$, we obtain the approximation of the circular variance in terms of $\kappa_{t}$:
\begin{align}
    \hat\uni_{t}\approx \rbra{1-\frac{1}{2}\kappa_{t}^{-1}}\uni(\theta_{t}^{\max}),\ 
    \mathrm{CV}[\theta_{t}-\hat\theta_{t}] \approx \frac{1}{2}\kappa_{t}^{-1}. \label{eq:cv_quadratic}
\end{align}
Eq. \eqref{eq:cv_quadratic} shows that the maximum point $\theta_{t}^{\max}$ approximately coincides with the estimated direction $\hat\theta_{t}:=\arg(\hat\uni_{t})$.
Therefore, the concentration $\kappa_{t}$ of the posterior density around the estimated direction $\hat\theta_{t}$ is sufficient for expressing the estimation limit.

We finally get the estimation limit by calculating the behavior of the observation and prediction terms around the estimated direction.
The prediction term around the estimated direction is characterized by $\kappa_{t}$:
\begin{align}
    \dif^{\pred}\log\Post_{t}(\theta) 
    &= \sbra{\ps{\log\Post_{t}(\theta)}{\theta} + \rbra{\pf{\log\Post_{t}(\theta)}{\theta}}^{2}}\dif t\\
    &\approx \kappa_{t}^{2}(\theta-\hat\theta_{t})^{2}\dif t. \label{eq:pred_quadratic}
\end{align}
The observation term (Eq. \eqref{eq:obs_fourier_final}) is expanded around $\theta=\hat\theta_{t}$ as follows:
\begin{align}
    \dif^{\sen}\log\Post_{t}(\theta)
    &\approx-\frac{1}{2}\ps{}{\theta}\dif^{\sen}\log\Post_{t}(\hat\theta_{t})\cdot(\theta-\hat\theta_{t})^{2}\\
    &\approx-\frac{1}{2}\Lambda\rbra{1-\frac{3}{2}\sigma\epsilon}(\theta-\hat\theta_{t})^{2}\dif t, \label{eq:obs_quadratic}
\end{align}
where we used the approximation $|\hat\uni_{t}|\approx 1$ which holds under large $\kappa$ with Eq. \eqref{eq:cv_quadratic}.
By combining Eqs. \eqref{eq:pred_quadratic} and \eqref{eq:obs_quadratic}, we obtain the stationary value of $\kappa_{t}$ satisfying $\dif\log\Post_{t}(\theta)=0$:
\begin{align}
    \kappa_{t} \approx \sqrt{\frac{\Lambda}{2}}\rbra{1-\frac{3}{4}\sigma\epsilon}. \label{eq:kappa_stationary}
\end{align}
We note that this relation between $\kappa_{t}$ and $\Lambda$ implies that the assumption of large $\kappa_{t}$ under large $\Lambda$ is self-consistent.
By substituting Eq. \eqref{eq:kappa_stationary} into Eq. \eqref{eq:cv_quadratic}, we obtain the approximate estimation limit at the stationary state:
\begin{align}
    \mathrm{CV}(\theta_{t}-\hat\theta_{t}) \approx \frac{1}{\sqrt{2\Lambda}}\rbra{1+\frac{3}{4}\sigma\epsilon}. \label{eq:estimation_limit}
\end{align}
This result indicates that the estimation is improved when $\sigma<0$, i.e., the cell elongates perpendicular to the estimated direction, and impaired when $\sigma>0$, i.e., the cell elongates along the estimated direction.
We note that this estimation limit reproduces previous result derived for a circular cell under a time-discrete setting when $\epsilon=0$ \cite{novak2021bayesian}.

\subsection{Numerical validation of estimation limit}
\begin{figure*}
    \centering
    \includegraphics[width=\textwidth]{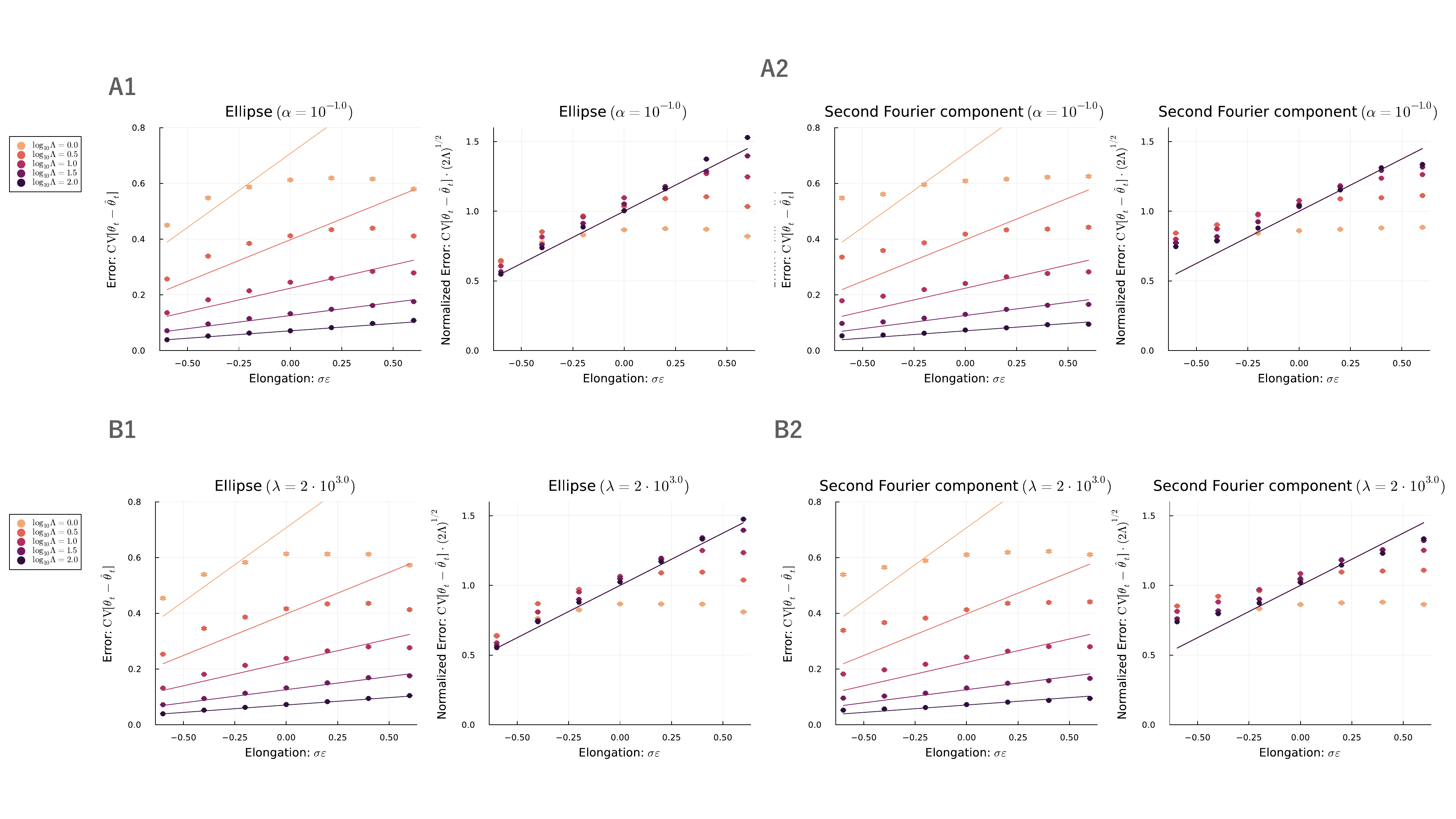}
    \caption{Estimation error and normalized estimation error as functions of the shape parameter $\sigma\epsilon$ for different values of the signal-to-noise ratio $\Lambda$. (A1, A2) Estimation error for an elliptic shape (A1) and a shape described by the second Fourier component (A2). Different curves represent different values of $\Lambda$, which are obtained by varying the ligand binding rate $\bar{\lambda}$ while keeping the gradient steepness $\alpha$ fixed. (B1, B2) Estimation error for an elliptic shape (B1) and a shape described by the second Fourier component (B2). Different curves represent different values of $\Lambda$, which are obtained by varying the gradient steepness $\alpha$ while keeping the ligand binding rate $\bar{\lambda}$ fixed. In all panels, solid lines represent theoretical predictions, while markers indicate simulation results.}
    \label{fig:estimation limits}
\end{figure*}

To validate the theoretical prediction that the control of elongation direction can improve or impair the estimation performance, as indicated by Eq. \eqref{eq:estimation_limit}, we perform numerical simulations of the gradient sensing process under different cell shape configurations (Figure \ref{fig:elongation_pattern}(B)).
We consider two types of cell shapes: a shape described by the second Fourier mode and an elliptic shape. The shape described by the second Fourier mode serves as a simple setting where only the second mode has non-zero values, and the elliptic shape is a widely used candidate for modeling elongating cells.

To ensure a fair comparison across elongation parameters, we rescale the radius as $r_{t}^{\phi} \rightarrow r_{t}^{\phi}/\sqrt{S(\epsilon)/S(0)}$, where $S(\epsilon)$ represents the area of the elongated shape with elongation parameter $\epsilon$, and $S(0)$ represents the area of the circular shape (i.e., when $\epsilon = 0$). This rescaling factor adjusts the size of the cell so that its area is independent of the elongation parameter $\epsilon$. The factor $S(\epsilon)/S(0)$ deviates from unity only by $o(\epsilon)$, and consequently, the rescaling does not change the analytical expression of the estimation limit given by Eq. (24).

In our simulations, the gradient direction $\theta_t$ evolves according to Eq. \eqref{eq:model_dimless_theta}, while the ligand-receptor binding is modeled as a Poisson process with the rate given by Eq. \eqref{eq:model_dimless_rate}. The posterior distribution of the gradient direction is computed by discretizing the filtering equation (Eq. \eqref{eq:filter}) over the direction $\theta$.

To quantify the estimation performance, we calculate the circular variance (Eq. \eqref{eq:circular variance}) by approximating the expectation using the Monte Carlo method with $10^4$ samples. The initial state of the posterior density function is set to be uniform. We observe that the circular variance reaches a nearly stationary value after a time period of $2\Lambda^{-1/2}$ (see Figure \ref{fig:error_dynamics}). Therefore, we define the stationary estimation limit as the average of the circular variance over the time interval $t \in [2\Lambda^{-1/2}, 3\Lambda^{-1/2}]$.

Figure \ref{fig:estimation limits} presents the estimation limit for various values of the signal-to-noise ratio $\Lambda$ and the elongation parameter $\sigma\epsilon$.
In the high signal-to-noise ratio regime, the results confirm our theoretical prediction: the estimation limit improves when the cell elongates perpendicular to the estimated direction ($\sigma<0$) and deteriorates when the cell elongates along the estimated direction ($\sigma>0$).
This trend holds for both the shape described by the second Fourier mode and the elliptic shape.
Moreover, the dependence of the estimation limit on the elongation magnitude $\epsilon$ is well captured by the analytical approximation (Eq. \eqref{eq:estimation_limit}) for both cell shapes.
This finding suggests that the second mode of the cell shape plays a dominant role in determining the estimation performance under linear concentration gradient.

Even when the signal-to-noise ratio decreases, the analytical formula explains the simulation results moderately well for $\Lambda \geq 10$. While simulation results start to deviate from the analytical prediction as $\Lambda$ decreases below 10, the qualitative result that the estimation improves for $\sigma < 0$ remains valid even for moderate values of $\Lambda = 1$, where estimation becomes increasingly difficult (see Figure \ref{fig:schematic_lowSNR} for sample trajectories of $\theta_{t}$ and $\hat\theta_{t}$).

In the ellipse case with $\Lambda=1$, a slight improvement in estimation is also observed for $\sigma >0$. This suggests the presence of additional behaviors that are not captured by our current analysis, which is reasonable given the qualitative differences in the low SNR regime.

Despite some deviations in the low SNR regime, these numerical results quantitatively support our theoretical analysis, demonstrating that the strategic control of cell shape elongation can indeed enhance or hinder the accuracy of gradient sensing, depending on the orientation of the elongation relative to the estimated gradient direction.
Further investigation into the low SNR regime may reveal additional insights and will be a subject for future work.

\section{Discussion}
\label{Discussion}
In this study, we utilized Bayesian nonlinear filtering theory to derive the estimation limits for fluctuating gradient directions under the feedback control of cell shape. By applying the diffusion approximation to the stochastic ligand-receptor binding process and the Laplace approximation for posterior expectation, we obtained an analytical expression for the estimation limit valid in the high signal-to-noise ratio (SNR) regime. Our results align with previously identified limits for circular cell shapes in discrete-time settings \cite{novak2021bayesian}.

Our key finding is that incorporating active cell shape control can enhance gradient sensing accuracy. We showed that when cells elongate perpendicular to the estimated gradient direction, the estimation error is reduced compared to the case of a static circular cell shape. This result parallels with previous theoretical findings that elongation perpendicular to the true gradient direction improves estimation performance \cite{hu2011geometry}. Our work extends this previous result by considering the more realistic scenario where the elongation direction is determined by sensory information and fluctuates around the true gradient direction due to sensory noise. We demonstrate that even in this case, dynamic shape control can enhance direction estimation.

An important future direction is to identify scenarios where cell shape dynamics provides biologically more significant enhancement of gradient sensing accuracy. While our focus on the high SNR regime allowed us to obtain the analytical expression, the observed improvement of estimation limit implicates only marginal enhancement in chemotaxis ability.
To clarify this point, we note that chemotaxis index, defined as the average migration along the gradient direction, $\mathrm{CI}=\Expect[\cos(\theta_{t}-\hat\theta_{t})]$, relates to the estimation error as $\mathrm{CI} = 1-\mathrm{CV}[\theta_{t}-\hat\theta_{t}]$, if we assume that cells migrate toward the estimated direction $\hat\theta_{t}$ and the estimated direction fluctuates symmetrically around the true gradient $\Expect[\sin(\theta_{t}-\hat\theta_{t})]=0$.
Therefore, the same amount of reduction in the estimation error is more pronounced under the lower SNR regime in terms of the increasing rate of the chemotaxis index.
We may find biologically significant improvement of gradient sensing ability by extensively investigating low SNR regime.

Another valuable extension of our analysis would be to account for uncertainty in the gradient steepness, especially in the low SNR regime.
In our current study, we implicitly assumed that the gradient steepness varies on a slower timescale than the gradient direction, allowing us to treat the steepness as a known constant. However, in reality, the gradient steepness can change significantly during migration, and previous work has shown that uncertainty in the steepness can have a notable impact on the maximum likelihood estimate of the gradient direction \cite{baba2012directional}.
Under the low SNR regime, the estimated direction is shown to be biased perpendicularly to or along with the elongation axis when the steepness is known or unknown, respectively.
Our Bayesian filtering framework provides a natural way to incorporate steepness uncertainty by modeling the steepness as a stochastic process with a characteristic fluctuation timescale. By tuning this timescale, we can interpolate the two extreme cases of known and unknown steepness. This approach could reveal interesting transitions in the bias of the estimated direction and, consequently, the optimal sensing strategy as a function of the steepness uncertainty. Characterizing such transitions and their dependence on SNR and other model parameters is an intriguing direction for future research.

\section*{Acknowledgements}
The first author received a JSPS Research Fellowship (Grant Number 20J21362). This research was supported by JSPS KAKENHI (Grant Number 19H05799 and 24H02148) and JST CREST (Grant Number JPMJCR2011).

\bibliographystyle{apsrev4-2}
\providecommand{\noopsort}[1]{}\providecommand{\singleletter}[1]{#1}

\appendix
\renewcommand{\thefigure}{\Alph{section}.\arabic{figure}}
\label{Appendix}

\section{Estimation limit of directional statistics}
\label{Appendix/Estimation_limit}
We derive the equations for estimation limit expressed with Eqs. \eqref{eq:estimation_optimal_estimator} and \eqref{eq:limit_min} by showing that the estimation error $\mathrm{CV}[\theta_{t}-\vartheta_{t}[\obs_{0:t}]]$ is bounded from below by $1-\Expect^{\obs_{0:t}}[\hat\uni_{t}]$ and the bound is attained by $\vartheta_{t}[\obs_{0:t}]=\hat\theta_{t}$.
To do that, we note that $\Expect^{\theta_{t}}[\exp(i\theta_{t})\mid\obs_{0:t}]$ on a complex plane can be idenfied with the real vector $\hat\uni_{t}:=\Expect^{\theta_{t}}[\uni(\theta_{t})\mid\obs_{0:t}]$ and thus $\Expect^{\theta_{t}}[\exp(\ii\theta_{t})\mid\obs_{0:t}] = \|\hat\uni_{t}\|\exp(\ii\hat\theta_{t})$ holds.
By using the property of conditional expectation, we can evaluate the estimation error as follows:
\begin{widetext}
\begin{align}
    \mathrm{CV}\sbra{\theta_{t}-\vartheta_{t}[\obs_{0:t}]} 
    &:=1-\left|\Expect^{\theta_{t},\obs_{0:t}}\sbra{\exp\rbra{\mathrm{i}(\theta_{t}-\vartheta_{t}[\obs_{0:t}])}}\right|\\
    &= 1-\left|\Expect^{\obs_{0:t}}\sbra{\Expect^{\theta_{t}}\sbra{\exp\rbra{\ii\theta_{t}}\mid\obs_{0:t}}\exp\rbra{-\ii\vartheta_{t}[\obs_{0:t}]}}\right|\\
    &= 1-\left|\Expect_{\obs}\sbra{\|\hat\uni_{t}\|\exp(\ii\hat\theta_{t})\exp\rbra{-\ii\vartheta_{t}[\obs_{0:t}]}}\right|\\
    &= 1-\sqrt{\Expect_{\obs}\sbra{\|\hat\uni_{t}\|\cos(\hat\theta_{t}-\vartheta_{t}[\obs_{0:t}])}^{2}+ \Expect_{\obs}\sbra{\|\hat\uni_{t}\|\sin(\hat\theta_{t}-\vartheta_{t}[\obs_{0:t}])}^{2}}\\
    &\geq 1-\sqrt{\Expect_{\obs}\sbra{\|\hat\uni_{t}\|}^{2}\rbra{\Expect_{\obs}\sbra{\cos(\hat\theta_{t}-\vartheta_{t}[\obs_{0:t}])}^{2}+\Expect_{\obs}\sbra{\sin(\hat\theta_{t}-\vartheta_{t}[\obs_{0:t}])}^{2}}}\\
    &\geq 1-\sqrt{\Expect_{\obs}\sbra{\|\hat\uni_{t}\|}^{2}\rbra{\Expect_{\obs}\sbra{\cos^{2}(\hat\theta_{t}-\vartheta_{t}[\obs_{0:t}])+\sin^{2}(\hat\theta_{t}-\vartheta_{t}[\obs_{0:t}])}}}\\
    &= 1-\Expect_{\obs}\sbra{\|\hat\uni_{t}\|}.
\end{align}
\end{widetext}
Here, we used the Cauchy-Schwartz inequality in the fifth line and the Jensen inequality in the six-th line.
We can check that $\hat\theta_{t}$ attains the lower bound by substituting $\vartheta_{t}[\obs_{0:t}]=\hat\theta_{t}$ in the third line.
Therefore, the optimal estimator and estimation limit are given by the formulae in Eq. \eqref{eq:estimation_optimal_estimator} and Eq. \eqref{eq:limit_min}, respectively.

\section{Filtering process marginalized over hidden process and diffusion approximation of Poisson observation term}
\label{Appendix/Filtering_general}
We explain how to obtain the dynamics of the posterior density $\Post_{t}$ averaged over the stochastic realization of true gradient $\theta_{t}$ dynamics and approximate the jump process $\dif\obs_{t}$ with a diffusion process.
Because the derivation can be applied not only to our setting but also to other filtering process, we explain the result based on a more general setting.

We formulate a filtering problem by modeling hidden and observation processes.
We consider a hidden process whose probability density function $\Prob_{t}(x) = \Prob(X_{t}=x)$ follows the Fokker-Planck equation
\begin{align}
    \pf{\Prob_{t}(x)}{t} = \mathcal{L}_{X}^{\ast}\Prob_{t}(x)
\end{align}
where $\mathcal{L}^{\ast}$ is the Fokker-Planck operator.
We model observation processes $\obs_{t}^{i}$ with $i\in\{1,\ldots,n_{\obs}\}$ by the Poisson processes whose rate processes are $\lambda_{t}^{i} = \lambda^{i}(X_{t},\Post_{t^{-}})$, i.e., $\Prob(\dif\obs_{t}^{i}\mid X_{t},\obs_{0:t^{-}}) = \lambda^{i}(X_{t},\Post_{t^{-}})\dif t$. Note that $\Post_{t^{-}}(x)=\Prob(X_{t^{-}}=x\mid \obs_{0:t^{-}})$ is a functional of $\obs_{0:t^{-}}.$
The posterior density function $\Post_{t}(x)$ under this setting follows the filtering equation (see \cite{kutschireiter2020hitchhiker} for pedagogical introduction of the filtering equation):
\begin{widetext}
\begin{align}
    \dif\Post_{t}(x) = \mathcal{L}_{X}^{\ast}\Post_{t}(x)\dif t + \Post_{t^{-}}(x)\sum_{j=1}^{N_{Y}}\frac{\lambda^{i}(x,\Post_{t^{-}})-\hat{\lambda}^{i}[\Post_{t^{-}}]}{\hat{\lambda}^{i}[\Post_{t^{-}}]}(\dif\obs_{t}^{i}-\hat{\lambda}^{i}[\Post_{t^{-}}]\dif t)\label{eq:dz_original}
\end{align}
\end{widetext}
where $\hat{f}[z]:= \int\dd x f(x,z)z(x)$ represents the posterior expectation of $f$ when the posterior density is $\Post_{t}=z$.
The problem in the main text corresponds to the case where the hidden process is $X_{t}=\theta_{t}$ with $\mathcal{L}^{\ast}_{X}\Post_{t}(X) = \ps{\Post_{t}}{X}(X)$ and the observation process is $\obs_{t}^{i} = \obs_{t}^{\phi_{i}}$ with $N_{\obs}=N$ and $\lambda^{i}(\theta,z) = (\bar\lambda/N)\cdot (1 + \alpha r^{\phi_{i}}[z]\abra{\uni(\phi_{i}),\uni(\theta)}$ where $r^{\phi_{i}}[z] = 1+\sigma\epsilon\cos(2(\phi-\hat\theta[z]))$ and $\hat\theta[z]$ is the posterior expectation of $\theta_{t}$ when the posterior density is $\Post_{t}=z$.

We then consider the time evolution of the joint density function of $X_{t}$ and $\Post_{t}$: $\Prob_{t}(x,\post):=\Prob(X_{t}=x,\Post_{t}=z)$.
Assuming that $\Post_{t}(\cdot)$ can be treated in the same way as a finite-dimensional stochastic variable, the dynamics of the joint density is formally represented by the following forward Kolmogorov equation:

\begin{widetext}
\begin{align}
    \pf{\Prob_{t}(x,\post)}{t} &= \mathcal{L}_{X}^{\ast}\Prob_{t}(x,\post) - \int\dd\tilde{x}\ff{}{z(\tilde{x})}\sbra{\mathcal{L}_{X}^{\ast}\post(\tilde{x})-\post(\tilde{x})\sum_{j=1}^{N_{Y}}(\lambda^{i}(\tilde{x},z)-\hat{\lambda}^{i}[\post])}\Prob_{t}(x,\post)\\
    &+ \int\dd w^{i} \cbra{\lambda_{\Post}^{i}(w^{i},x,\post-w^{i})\Prob_{t}(x,\post-w^{i}) - \lambda_{\Post}^{i}(w^{i},x,\post)\Prob_{t}(x,\post)}
\end{align}
\end{widetext}
where $\lambda_{\Post}^{i}(w,x,\post) := \lambda^{i}(x,z)\delta\rbra{w^{i}(\cdot) - \post(\cdot)\frac{\lambda^{i}(\cdot,z)-\hat{\lambda}^{i}[\post]}{\hat{\lambda}^{i}[\post]}}$ describes the rate of jump of $\post$ with size $w$ when $X_{t}=x$ and  $\Post_{t}=\post$, and  $\delta/\delta z(\tilde{x})$ represents a functional derivative, which describes how a functional changes with respect to variations in its input function $z$ at the point $\tilde{x}$ (see Ref. \cite{kanazawa2020field}, Sec. II.C.5, for the treatment of functional derivatives in the context of stochastic processes).
We note that there is a distinction between $x$ and $\tilde{x}$ which represent the realization of the hidden process $X_{t}$ and the argument of the posterior density $Z_{t}$, respectively.

To marginalize over hidden variable $X_{t}$, we integrate each term with respect to $x$ by noting that $\Prob_{t}(x,z)=\Prob_{t}(x\mid z)\Prob_{t}(z)=z(x)\Prob_{t}(z)$:
\begin{widetext}
\begin{align}
    \pf{\Prob_{t}(\post)}{t} &= -\int\dd\tilde{x}\ff{}{z(\tilde{x})}\sbra{\mathcal{L}_{X}^{\ast}\post(\tilde{x})\Prob_{t}(\post)-\post(\tilde{x})\sum_{j=1}^{N_{Y}}(\lambda^{i}(\tilde{x},z)-\hat{\lambda}^{i}[\post])\Prob_{t}(\post)}\\
    &+\sum_{j=1}^{N_{Y}}\cbra{\int\dd w^{i} \cbra{\hat{\lambda}_{\Post}^{i}[w^{i},\post-w^{i}]\Prob_{t}(\post-w^{i}) - \hat{\lambda}_{\Post}^{i}[w^{i},\post]\Prob_{t}(\post)}}, \label{eq:dP(z)_jump}
\end{align}
\end{widetext}
where $\hat{\lambda}_{\Post}^{i}[w,\post] := \hat{\lambda}^{i}[\post]\delta\rbra{w^{i}(\cdot) - \post(\cdot)\frac{\lambda^{i}(\cdot,z)-\hat{\lambda}^{i}[\post]}{\hat{\lambda}^{i}[\post]}}$ represents the posterior expectation of $\lambda_{Z}^{i}(w,x,z)$.

Next, we approximate the jump term by a diffusion based on the Kramers-Moyal expansion.
To do that, we add an assumption on the observation process.
The jump rate $\lambda^{i}$ in the observation process can be decomposed into a sum of $X_{t}$-dependent and $X_{t}$-independent terms: $\lambda^{i}(x,z) = \bar\lambda^{i}(1+\alpha^{i}(x,z))$.
For the $X_{t}$ dependent term, we assume that $\alpha^{i}(x,z)$ is sufficiently small so that $\alpha^{i}(x,z)=O(\alpha)$ holds for any $j$ and $x$ with a small parameter $\alpha$ and that terms of order $o(\alpha^{2})$ are ignorable.
Then, we expand the jump term in the Taylor series about the jump size $w^{i}$:
\begin{widetext}
\begin{align}
    \int\dd w^{i} \cbra{\hat{\lambda}_{\Post}^{i}[w^{i},\post-w^{i}]\Prob_{t}(\post-w^{i}) - \hat{\lambda}_{\Post}^{i}[w^{i},\post]\Prob_{t}(\post)}
    &= \int\dd w^{i} \int\dd\tilde{x}^{\mi}\sum_{|\mi|=1}^{\infty}\frac{(-w^{i}(\tilde{x}))^{\mi}}{|\mi|!}\frac{\delta^{\mi}}{\delta\post(\tilde{x})^{\mi}}\sbra{\hat{\lambda}_{\Post}^{i}[w^{i},\post]\Prob_{t}(\post)}\\
    &= \int\dd\tilde{x}^{\mi}\sum_{|\mi|=1}^{\infty}\frac{(-1)^{|\mi|}}{|\mi|!}\frac{\delta^{\mi}}{\delta\post(\tilde{x})^{\mi}}\sbra{\int\dd w^{i}w^{i}(\tilde{x})^{\mi}\hat{\lambda}_{\Post}^{i}[w^{i},\post]\Prob_{t}(\post)}\\
    &=\int\dd\tilde{x}^{\mi}\sum_{|\mi|=1}^{\infty}\frac{(-1)^{|\mi|}}{|\mi|!}\frac{\delta^{\mi}}{\delta\post(\tilde{x})^{\mi}}\sbra{\hat{\lambda}^{i}[\post]\rbra{\post(\tilde{x})\frac{\lambda^{i}(\tilde{x},z)-\hat{\lambda}^{i}[\post]}{\hat{\lambda}^{i}[\post]}}^{\mi}\Prob_{t}(\post)} \label{eq:Kramers-Moyal}
\end{align}
\end{widetext}
where $\mi$ is a multi-index.
By noting that $\rbra{\post(\tilde{x})\frac{\lambda^{i}(\tilde{x},z)-\hat{\lambda}^{i}[\post]}{\hat{\lambda}^{i}[\post]}}^{\mi}$ is of the order $O(\alpha^{|\mi|})$, we ignore the third and higher order terms with respect to $|\mi|$.
Because the first order term of \eqnref{eq:Kramers-Moyal} cancels with the term $-\sum_{j=1}^{N_{Y}}(\lambda^{i}(\tilde{x},z)-\hat{\lambda}^{i}[\post])\Prob_{t}(\post)$ in the first line of \eqnref{eq:dP(z)_jump}, we obtain the dynamics of $\Prob_{t}(z)$ as the following Fokker-Planck equation:
\begin{widetext}
\begin{align}
    \pf{\Prob_{t}(\post)}{t} &= -\int\dd\tilde{x}\ff{}{z(\tilde{x})}\sbra{\mathcal{L}_{X}^{\ast}\post(\tilde{x})\Prob_{t}(\post)}\\
    &\quad+\frac{1}{2}\sum_{j=1}^{N_{Y}}\int\dd\tilde{x}\dd\tilde{x}'\fss{}{\post(\tilde{x})}{\post(\tilde{x}')}\sbra{\frac{\post(\tilde{x})(\lambda^{i}(\tilde{x},z)-\hat{\lambda}^{i}[\post])\cdot\post(\tilde{x}')(\lambda^{i}(\tilde{x}',z)-\hat{\lambda}^{i}[\post])}{\hat\lambda^{i}[\post]}\Prob_{t}(\post)}\\
    &= -\int\dd\tilde{x}\ff{}{z(\tilde{x})}\sbra{\mathcal{L}_{X}^{\ast}\post(\tilde{x})\Prob_{t}(\post)}\\
    &\quad+\frac{1}{2}\sum_{j=1}^{N_{Y}}\int\dd\tilde{x}\dd\tilde{x}'\fss{}{\post(\tilde{x})}{\post(\tilde{x}')}\bar\lambda^{i}\sbra{\post(\tilde{x})(\alpha^{i}(\tilde{x},z)-\hat{\alpha}^{i}[\post])\cdot\post(\tilde{x}')(\alpha^{i}(\tilde{x},z')-\hat{\alpha}^{i}[\post])\Prob_{t}(\post)}
\end{align}
\end{widetext}
This Fokker-Planck equation describes the following SDE of $\Post_{t}$:
\begin{widetext}
\begin{align}
    \dif\Post_{t}(x)
    &= \mathcal{L}_{X}^{\ast}\Post_{t}(x)\dif t + \Post_{t}(x)
    \sum_{j=1}^{N_{Y}}\sqrt{\bar\lambda}\cbra{\alpha^{i}(x,z)-\hat{\alpha}^{i}[\Post_{t}]}\dif\W_{t}^{i} \label{eq:dz_general}
\end{align}
\end{widetext}
where $\dif\W_{t}^{i}$ are the standard Wiener processes.
By applying the result to our problem in the main text, the prediction term does not change from Eq. \eqref{eq:filter_pred} and the observation process is transformed from Eq. \eqref{eq:filter_obs} to Eq. \eqref{eq:dz_general}.

\section{Relaxation of the estimation error to the stationary state}
\label{Appendix/Error_dynamics}
\begin{figure*}
    \centering
    \includegraphics[width=\linewidth]{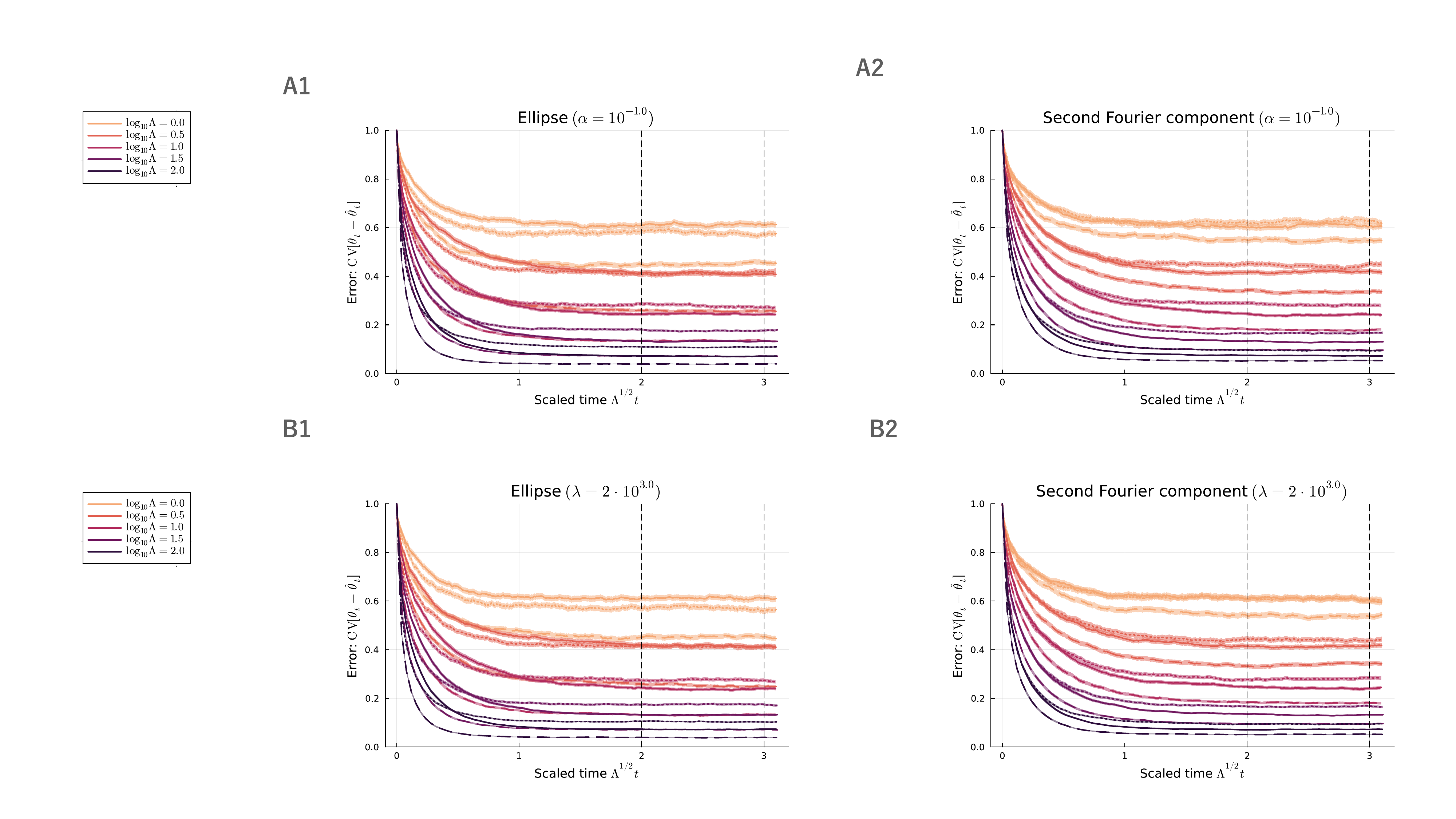}
    \caption{Estimation error as a function of the scaled time $\Lambda^{1/2}t$ for different values of the signal-to-noise ratio $\Lambda$ and the shape parameter $\sigma\epsilon$. (A1, A2) Estimation error for an elliptic shape (A1) and a shape described by the second Fourier component (A2). Different curves represent different values of $\sigma\epsilon$ and $\Lambda$, obtained by varying the ligand binding rate $\bar{\lambda}$ while keeping the gradient steepness $\alpha$ fixed. (B1, B2) Estimation error for an elliptic shape (B1) and a shape described by the second Fourier component (B2). Different curves represent different values of $\epsilon$ and $\Lambda$, obtained by varying the gradient steepness $\alpha$ while keeping the ligand binding rate $\bar{\lambda}$ fixed. In all panels, solid, dotted, and dashed curves represent $\sigma\epsilon=0$, $\sigma\epsilon=0.7$, and $\sigma\epsilon=-0.7$, respectively. The shaded regions around curves indicate $\pm$ the standard error estimated by bootstrap sampling. The vertical dashed lines indicate the time interval used for time averaging, from $t=2\Lambda^{-1/2}$ to $t=3\Lambda^{-1/2}$.}
    \label{fig:error_dynamics}
\end{figure*}

To show that the estimation error reaches a stationary state after time depending on the parameter $\Lambda$, we plot the estimation error as a function of the scaled time $\Lambda^{1/2}t$ in Figure \ref{fig:error_dynamics}.

\section{Sample trajectories of the filtering processes under low signal-to-noise ratio regime}
\label{Appendix/Filtering_trajectory}
\begin{figure*}
    \centering
    \includegraphics[width=\linewidth]{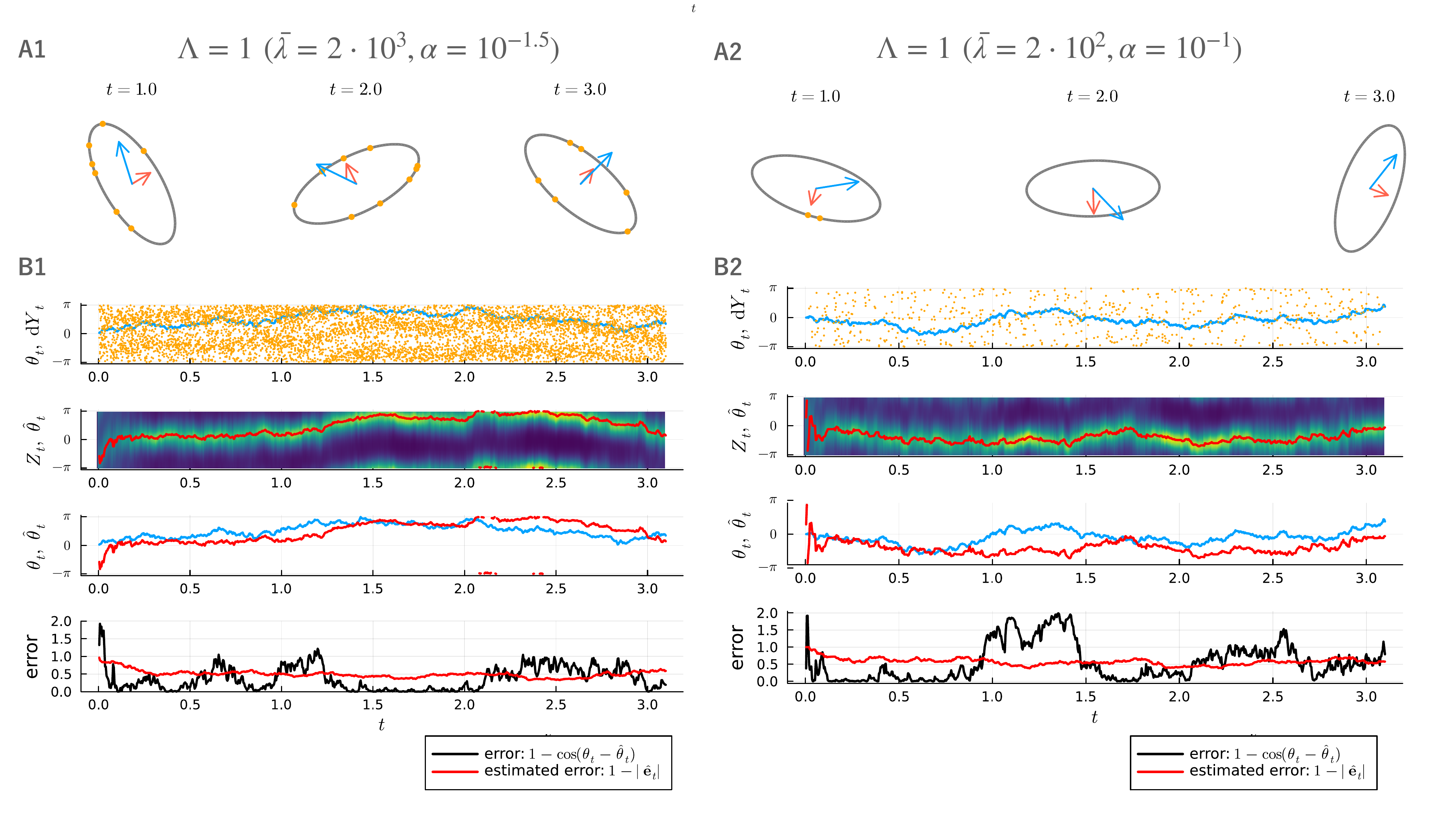}
    \caption{
    Gradient sensing processes under a signal-to-noise ratio lower than that in Fig. \ref{fig:schematic}.
    (A1,A2) Time lapse of a sample process of the gradient sensing when the signal-to-noise ratio is decreased to $\Lambda=1$ by reducing $\bar\lambda$ from $2\cdot 10^3$ to $2\cdot 10^2$ (A1) and by reducing $\alpha$ from $10^{-1}$ to $10^{-1.5}$ (A2).
    Blue and red arrows indicate the true and estimated gradient direction, respectively.
    The length of the red arrows indicate the confidence of estimate $|\hat\uni_{t}|$.
    Orange dots show the locations of receptors at which ligand bound during time $[t,t+\Delta t)$ with $\Delta t$ being the discretization width.
    (B1,B2)Time evolution of variables in the sample process shown in (A1,A2) when the signal-to-noise ratio is decreased to $\Lambda=1$ by reducing $\bar\lambda$ from $2\cdot 10^3$ to $2\cdot 10^2$ (A1) and by reducing $\alpha$ from $10^{-1}$ to $10^{-1.5}$ (A2).
    The first row shows the true gradient direction $\theta_{t}$ (blue curve) and the angular locations of receptors at which ligand binds $\dif\obs_{t}$ (orange dots).
    The second row shows the posterior density of the gradient direction $\Post_{t}$ and the estimated gradient direction $\hat\theta_{t}$ (red line).
    The third row shows the true (blue curve) and estimated (red curve) gradient direction.
    The final row shows the error between the true and estimated direction measured by $(\theta_{t}-\hat\theta_{t})^{2}/2$ (black curve) and the uncertainty of estimate represented by $1-|\hat\uni_{t}|$ (red curve).
    The other parameter is set to $\sigma\epsilon = -0.4$ during the simulation.}
    \label{fig:schematic_lowSNR}
\end{figure*}

To complement the parameter dependence of the estimation error, Figure \ref{fig:schematic_lowSNR} shows the time trajectory of the true and estimated gradient direction with the signal-to-noise ratio being lower than the value used in Figure \ref{fig:schematic}.



\end{document}